\documentclass[conference]{IEEEtran}
\IEEEoverridecommandlockouts
\usepackage[2023, cc-by, pages={13}{18}, url=https://osda.ws/r/1njIu]{osda}
\usepackage{hyperref}%
\hypersetup{%
	colorlinks = true,
	allcolors  = black,
	breaklinks=false,   
	pdfusetitle=true,  
}%
\usepackage{cite}
\usepackage{amsmath,amssymb,amsfonts}

\usepackage{algorithmic}
\usepackage{graphicx}
\usepackage{textcomp}
\usepackage{xcolor}
\usepackage{colortbl}
\usepackage{tabularx}
\usepackage{multirow}
\usepackage{tikz}
\def\boxe[#1]{%
	\protect\tikz[#1]\protect\node [draw=black,fill=#1,thick,rectangle]{};%
}%

\usetikzlibrary{positioning}
\usetikzlibrary{shapes.multipart,shapes.geometric, arrows}
\usetikzlibrary{shapes.geometric, arrows, shadows,positioning}
\usepackage{subfig}

\usepackage{listings}
\definecolor{mygreen}{rgb}{0.2,0.6,0.2}
\definecolor{mygray}{rgb}{0.5,0.5,0.5}
\definecolor{mymauve}{rgb}{0.58,0,0.82}
\usepackage{flushend}
\usepackage{cleveref}

\usepackage[acronym, hyperfirst=false]{glossaries} 
\glsdisablehyper
\newacronym{xpe}{XPE}{Xilinx Power Estimator}
\newacronym{lfsr}{LFSR}{linear-feedback shift register}
\newacronym{lut}{LUT}{lookup table}
\newacronym{dut}{DUT}{device under test}
\newacronym{vcd}{VCD}{value change dump}
\newacronym{iverilog}{iverilog}{Icarus Verilog}

\newacronym{asd}{ASD}{Autism Spectrum Disorder}
\newacronym{cart}{CART}{Classification And Regression Tree}
\newacronym{dfa}{DFA}{Discriminant Function Analysis} 
\newacronym{ecg}{ECG}{Electrocardiography}
\newacronym{skin_eda}{EDA}{Electrodermal activity} 		
\newacronym{ews}{EWS}{Early Warning Score}
\newacronym{ga}{GA}{Genetic Algorithm}
\newacronym{hr}{HR}{Heart Rate}
\newacronym{ldf}{LDF}{Linear Discriminant Function}
\newacronym{rmsdd}{RMSDD}{Root-Mean Square of Successive Differences}
\newacronym{rsvm}{RSVM}{Reputation-driven Support Vector Machine }
\newacronym{saews}{SA-EWS}{Self-Aware Early Warning Score}
\newacronym{sam}{SAM}{Self-Assessment Manikin} 
\newacronym{selphys}{SelPhyS}{Self-aware cyber-Physical System}
\newacronym{sdnn}{SDNN}{Standard Deviation Normal-to-Normal-Intervals}
\newacronym{skt}{SKT}{Skin Temperature}
\newacronym{som}{SOM}{Self-Organizing Map}
\newacronym{whs}{WHS}{Wearable Health-care Systems}

\newacronym{ai}{AI}{Artificial Intelligence}
\newacronym{ann}{ANN}{Artificial Neural Network}
\newacronym{bpn}{BPN}{Back-Propagation Neural Network}
\newacronym{bpnn}{BPNN}{back-propagation neural network}
\newacronym{cnn}{CNN}{Convolutional Neural Network}
\newacronym{flop}{FLOP}{Floating Point Operation}
\newacronym{icnn}{ICNN}{Iterative Convolutional Neural Network}
\newacronym{svm}{SVM}{Support Vector Machine}
\newacronym{nn}{NN}{Neural Network}
\newacronym{nb}{NB}{Naive Bayesian}
\newacronym{mcs}{MCS}{Multiple Classifier System}
\newacronym{ml}{ML}{Machine Learning}
\newacronym{ucnn}{$\mu$CNN}{Micro CNN}

\newacronym{brs}{BRS}{Bipolar Resistive Switch-based logic}
\newacronym{cnf}{CNF}{Conjunctive Normal Form}
\newacronym{crs}{CRS}{Complementary Resistive Switch-based logic}
\newacronym{dnf}{DNF}{Disjunctive Normal Form}
\newacronym{ecm}{ECM}{Electrochemical Metallization}
\newacronym{felix}{FELIX}{Fast and Energy-Efficient Logic in Memory}
\newacronym{fpm}{FPM}{Forward Polarized Memristor}
\newacronym{hfo}{$HfO_x$}{Hafnium Oxide}
\newacronym{hrs}{HRS}{High Resistance State}
\newacronym{imc}{IMC}{In-Memory Computation}
\newacronym{imply}{IMPLY}{Material Implication}
\newacronym{nimp}{NIMP}{NOT IMPLY}
\newacronym{imp}{IMP}{In-Memory Processing}
\newacronym{lim}{LIM}{Logic in Memory}
\newacronym{lrs}{LRS}{Low Resistance State}
\newacronym{ltg}{LTG}{Logic Threshold Gate}
\newacronym{magic}{MAGIC}{Memristor-Aided Logic}
\newacronym{mecoins}{Me-Coin}{Memristor-based Computation In-memory}
\newacronym{pcm}{PCM}{Phase Change Memory}
\newacronym{pim}{PIM}{Processing in Memory}
\newacronym{reram}{ReRAM}{Resistive Random Access Memory}
\newacronym{rpm}{RPM}{Reversely Polarized Memristor}
\newacronym{stt}{STT}{Spin Transfer Torque}
\newacronym{tao}{$TaO_x$}{Tantalum Oxide}
\newacronym{tio}{$TiO_2$}{Titanium dioxide}
\newacronym{vc}{VC}{Valence Change}

\newacronym{aco}{ACO}{Autonomous Cooperating Object}
\newacronym{afdd}{AFDD}{Automated Fault Detection and Diagnostic}
\newacronym{ca}{CA}{Continuous Average}
\newacronym{cah}{CAH}{Context-Aware Health Monitoring}
\newacronym{cam}{CCAM}{Confidence-based Context-Aware condition Monitoring}
\newacronym[plural=DABs,longplural={Discrete Average Blocks}]{dab}{DAB}{Discrete Average Block}
\newacronym[plural=KPNs,longplural={Kahn Process Networks}]{kpn}{KPN}{Kahn Process Networks} 
\newacronym{mape-k}{MAPE-K}{Monitor-Analyze-Plan-Execute over a shared Knowledge}
\newacronym[plural=MoCs,longplural={Models of Computation}]{moc}{MoC}{Model of Computation}
\newacronym{oda}{ODA}{Observe-Decide-Act}
\newacronym{pca}{PCA}{Principal Component Analysis}
\newacronym{rosa}{RoSA}{Research on Self-Awareness}
\newacronym{sa}{SA}{Self-Aware}
\newacronym{saness}{SA}{Self-Awareness}
\newacronym{sahm}{SAHM}{Self-Aware Health Monitoring}
\newacronym{samba}{SAMBA}{Self-Aware health Monitoring and Bio-inspired coordination for distributed Automation systems}
\newacronym{sh}{SH}{State Handler}

\newacronym{cps}{CPS}{Cyber-Physical System}
\newacronym{cpps}{CPPS}{Cyber-Physical Production System}
\newacronym{dsr}{DSR}{Down-Sampling Rate}
\newacronym{dum}{DuM}{Device under Monitoring}
\newacronym{es}{ES}{Embedded System}
\newacronym{mes}{MES}{Manufacturing Execution System}
\newacronym{sos}{SoS}{System of Systems}
\newacronym{suo}{SuO}{System under Observation}

\newacronym{abi}{ABI}{Application Binary Interface}
\newacronym{adc}{ADC}{Analog-Digital Converter}
\newacronym{aes}{AES}{Advanced Encryption Standard}
\newacronym{alu}{ALU}{Arithmetic Logic Unit}
\newacronym{api}{API}{Application Programming Interface}
\newacronym{asic}{ASIC}{Application Specific Integrated Circuit}
\newacronym{asoc}{ASOC}{Autonomic System-on-Chip platform}
\newacronym{axi}{AXI}{Advanced eXtensible Interface Bus}
\newacronym{bram}{BRAM}{Block Random Access Memory}
\newacronym{cdt}{CDT}{C/C++ Development Tooling}
\newacronym{clb}{CLB}{Configuarable Logic Block}
\newacronym{cmos}{CMOS}{Complementary Metal-Oxide Semiconductor}
\newacronym{cp}{CP}{Clock Pulse}
\newacronym{cpi}{CPI}{Cycles Per Instruction}
\newacronym{cpu}{CPU}{Central Processing Unit} 
\newacronym{cpsoc}{CPSoC}{Cyber-Physical System-on-Chip}
\newacronym{cu}{CU}{Compute Unit}
\newacronym{cuda}{CUDA}{Compute Unified Device Architecture}
\newacronym{dac}{DAC}{Digital to Analog Converter}
\newacronym{ddr3}{DDR3}{Double Data Rate}
\newacronym{dff}{DFF}{Data Flip-Flop}
\newacronym{dll}{DLL}{Delay Locked Loop}
\newacronym{dmr}{DMR}{Dual Modular Redundancy}
\newacronym{dram}{DRAM}{Dynamic Random Access Memory}
\newacronym{dsd}{DSD}{Digital Synchronous Detection}
\newacronym{dsp}{DSP}{Digital Signal Processor}
\newacronym{dt}{DigiTime}{}
\newacronym{dvfs}{DVFS}{Dynamic Voltage and Frequency Scaling}
\newacronym{eda}{EDA}{Electronic Design Automation}
\newacronym{fdc}{FDC}{Frequency-to-Digital Converter}
\newacronym{fifo}{FIFO}{First In First Out}
\newacronym{fpga}{FPGA}{Field Programmable Gate Array}
\newacronym{gds}{GDS}{Global Data Share}
\newacronym{gnulgpl}{GNU LGPL}{GNU Lesser General Public Licence} 
\newacronym{gpgpu}{GPGPU}{General Purpose Graphics Processing Unit}
\newacronym{gpr}{GPR}{General Purpose Register}
\newacronym{gpu}{GPU}{Graphics Processing Unit}
\newacronym{gro}{GRO}{Gated Ring Oscillator}
\newacronym{io}{IO}{Input-Output}
\newacronym{hamsoc}{HAMSoC}{Hierarchical Agent Monitoring System-on-Chip}
\newacronym{hdl}{HDL}{Hardware Description Language}
\newacronym{hmp}{HMP}{Heterogeneous Multi-Processor}
\newacronym{ic}{IC}{Integrated Circuit}
\newacronym{icap}{ICAP}{Internal Configuration Access Port}
\newacronym[longplural={Intellectual Properties}]{ip}{IP}{Intellectual Property}
\newacronym{isa}{ISA}{Instruction Set Architecture}
\newacronym{lds}{LDS}{Local Data Share}
\newacronym{lru}{LRU}{Least Recently Used}
\newacronym{lsb}{LSB}{Least-Significant Bit}
\newacronym{lsu}{LSU}{Load Store Unit}
\newacronym{mash}{MASH}{Multi-Stage Noise-Shaping}
\newacronym{mems}{MEMS}{Micro-Electro-Mechanical Systems}
\newacronym{miaow}{MIAOW}{Many-core Integrated Accelerator Of deepwater/Wisconsin}
\newacronym{mosfet}{MOSFET}{Metal Oxide Semiconductor Field Effect Transistor}
\newacronym{mpsoc}{MPSoC}{Multi-Processor System-on-Chip}
\newacronym{mshr}{MSHR}{Miss Status Holding/Handling Register}
\newacronym{noc}{NoC}{Network-on-Chip}
\newacronym{opencl}{OpenCL}{Open Computing Language}
\newacronym{ocn}{OCN}{On-Chip Network}
\newacronym{pcb}{PCB}{Printed Circuit Board}
\newacronym{pcie}{PCIe}{Peripheral Component Interconnect Express}
\newacronym{pl}{PL}{Programmable Logic}
\newacronym{pli}{PLI}{Verilog Programming Language Interface}
\newacronym{pll}{PLL}{Phase-Locked Loop}
\newacronym{ps}{PS}{Processing System}
\newacronym{pv}{PV}{Process Variation}
\newacronym{qoe}{QoE}{Quality of Experience}
\newacronym{qos}{QoS}{Quality of Service}
\newacronym{ram}{RAM}{Random Access Memory} 
\newacronym{risc}{RISC}{Reduced Instruction Set Computer}
\newacronym{riscv}{RISC-V}{Reduced Instruction Set Computing - V}
\newacronym{rtl}{RTL}{Register-Transfer Level}
\newacronym{sdk}{SDK}{Software Development Kit}
\newacronym{seec}{SEEC}{SElf-awarE Computing}
\newacronym{sgpr}{SGPR}{Scalar General Purpose Register}
\newacronym{si}{SI}{Southern Island}
\newacronym{simd}{SIMD}{Single Instruction Multiple Data}
\newacronym{simf}{SIMF}{Single Instruction Multiple Floating point}
\newacronym{sm}{SM}{Streaming Multiprocessor}
\newacronym{snr}{SNR}{Signal to Noise Ratio}
\newacronym[plural=SoCs,firstplural=Systems on Chip (SoCs)]{soc}{SoC}{System-on-Chip}
\newacronym{spared}{SPARED}{Self-aware PArtial Reconfiguration architecture for Edge Devices}
\newacronym{spice}{SPICE}{Simulation Program With Integrated Circuit Emphasis}
\newacronym{tad}{TAD}{Time \gls{adc}}
\newacronym[plural=TDCs,longplural={Time-to-Digital Converters}]{tdc}{TDC}{Time-to-Digital Converter}
\newacronym{tq}{TQ}{Time-Quantizer}
\newacronym{uart}{UART}{Universal Asynchronous Receiver/Transmitter}
\newacronym{vcdu}{VCDU}{Voltage Controlled Delay Unit}
\newacronym{vco}{VCO}{Voltage Controlled Oscillator}
\newacronym{vga}{VGA}{Video Graphics Array}
\newacronym{vhdl}{VHDL}{Very High Speed Integrated Circuit Hardware Description Language}
\newacronym{vlsi}{VLSI}{Very Large Scale Integration}
\newacronym{vgpr}{VGPR}{Vector General Purpose Register}
\newacronym{xilffs}{XILFFS}{Generic Fat File System Library}

\newacronym{amd}{AMD}{Advanced Micro Devices}
\newacronym{beol}{BEOL}{Back End Of Line}
\newacronym{cad}{CAD}{Computer-Aided Design}
\newacronym{cas}{CAS}{Circuits and Systems}
\newacronym{eu}{EU}{European Union}
\newacronym{fdd}{FDD}{fault detection and diagnostic}
\newacronym{fefet}{FeFET}{Ferroelectric Field Effect Transistor}
\newacronym{feline}{FeLINe}{FeFET Logic IN mEmory}
\newacronym[plural=FOMs,longplural={Figures of Merit}]{fom}{FoM}{Figure of Merit}
\newacronym{hipeac}{HiPEAC}{High Performance and Embedded Architecture and Compilation}
\newacronym{hp}{HP}{Hewlett Packard}
\newacronym{hqp}{HQP}{Highly Qualified People}
\newacronym{hvac}{HVAC}{Heating, Ventilation and Air Conditioning}
\newacronym{ibm}{IBM}{International Business Machines corporation}
\newacronym{ict}{ICT}{Institute for Computer Technology}
\newacronym{iot}{IoT}{Internet of Things}
\newacronym{itrs}{ITRS}{International Technology Roadmap for Semiconductors}
\newacronym{irds}{IRDS}{International Roadmap for Devices and Systems}
\newacronym{nda}{NDA}{Non-Disclosure Agreement}
\newacronym{nvp}{NVP}{Non-Volatile Processor}
\newacronym{oecd}{OECD}{Organization for Economic Cooperation and Development}
\newacronym{rd}{R\&D}{Research and Development}
\newacronym{soa}{SoA}{State-of-the-Art}
\newacronym{tsmc}{TSMC}{Taiwan Semiconductor Manufacturing Company}
\newacronym{tvlsi}{TVLSI}{Transactions on Very Large Scale Integration}
\newacronym{vna}{VNA}{Von-Neumann Architecture}

\newacronym{fpu}{FPU}{Floating Point Unit}
\newacronym{i2c}{I2C}{Inter-Integrated Circuit}
\newacronym{spi}{SPI}{Serial Peripheral Interface}
\newacronym{gcc}{GCC}{GNU Compiler Collection}
\newacronym{udre}{UDRE}{UART Data Register Empty}
\newacronym{gps}{GPS}{Global Positioning System}
\newacronym{pc}{PC}{Pin-Change}
\newacronym{isr}{ISR}{Interrupt-Service-Routine}

\def\BibTeX{{\rm B\kern-.05em{\sc i\kern-.025em b}\kern-.08em
    T\kern-.1667em\lower.7ex\hbox{E}\kern-.125emX}}
\let\oldhref\href
\renewcommand{\href}[2]{\oldhref{#1}{\hbox{#2}}}
\begin{document}

\title{Towards Power Characterization of FPGA Architectures To Enable
Open-Source Power Estimation Using Micro-Benchmarks}

\author{\IEEEauthorblockN{Stefan Riesenberger}
\IEEEauthorblockA{\textit{TU Wien} \\
\textit{Institute of Computer Technology}\\
Vienna, Austria \\
stefan.riesenberger@alumni.tuwien.ac.at}
\and
\IEEEauthorblockN{Christian Krieg}
\IEEEauthorblockA{\textit{TU Wien} \\
\textit{Institute of Computer Technology}\\
Vienna, Austria \\
christian.krieg@alumni.tuwien.ac.at}
}
\glsdisablehyper
\maketitle
\begin{abstract}
While in the past decade there has been significant progress in open-source
synthesis and verification tools and flows, one piece is still missing in the
open-source design automation ecosystem: a tool to estimate the power
consumption of a design on specific target technologies. We discuss a
work-in-progress method to characterize target technologies using generic
micro-benchmarks, whose results can be used to establish power models of these
target technologies.  These models can further be used to predict the power
consumption of a design in a given use case scenario (which is currently out of
scope). We demonstrate our characterization method on the publicly documented
\emph{Lattice} iCE40 FPGA technology, and discuss two approaches to generating
micro-benchmarks which consume power in the target device: simple lookup table
(LUT) instantiation, and a more sophisticated instantiation of ring oscillators.
We study three approaches to stimulate the implemented micro-benchmarks in
hardware: Verilog testbenches, micro-controller testbenches, and pseudo-random
linear-feedback-shift-register-(LFSR)-based testing. We measure the power
consumption of the stimulated target devices. Our ultimate goal is to
automate power measurements for technology characterization; Currently, we
manually measure the consumed power at three shunt resistors using an
oscilloscope. Preliminary results indicate that we are able to induce variable
power consumption in target devices; However, the sensitivity of the power
characterization is still too low to build expressive power estimation models.

\end{abstract}

\begin{IEEEkeywords}
FPGA, power measurements, open source, yosys, nextpnr, iverilog
\end{IEEEkeywords}

\section{Introduction}

\begin{table}[tb]
    \footnotesize
    \renewcommand{\arraystretch}{1.2}
    \setlength\tabcolsep{3pt}%
    \centering
    \begin{tabular}{r|c|c|c|c|c|c|c|c|c|}
\multirow{3}{*}{\shortstack[l]{\boxe[black]\dots Full-Support\\\boxe[gray]\dots Partial-Support\\\boxe[white]\dots No-Support}}&\multicolumn{9}{c|}{}\\
&\multicolumn{9}{c|}{Power}\\
& \rotatebox{90}{General}& \rotatebox{90}{Per Voltage Source}& \rotatebox{90}{Per General Block}& \rotatebox{90}{Per Element}& \rotatebox{90}{Per Signal}& \rotatebox{90}{Per Data}& \rotatebox{90}{General IO}& \rotatebox{90}{Per IO}& \rotatebox{90}{Clock Domain}\\\hline
Lattice iCEcube2&\cellcolor{black}&\cellcolor{white}&\cellcolor{white}&\cellcolor{white}&\cellcolor{white}&\cellcolor{white}&\cellcolor{black}&\cellcolor{white}&\cellcolor{black}\\\hline
Lattice Diamond&\cellcolor{black}&\cellcolor{black}&\cellcolor{black}&\cellcolor{white}&\cellcolor{white}&\cellcolor{white}&\cellcolor{black}&\cellcolor{black}&\cellcolor{black}\\\hline
Intel PowerPlay&\cellcolor{black}&\cellcolor{black}&\cellcolor{black}&\cellcolor{white}&\cellcolor{black}&\cellcolor{white}&\cellcolor{white}&\cellcolor{white}&\cellcolor{black}\\\hline
Xilinx Power Estimator&\cellcolor{black}&\cellcolor{black}&\cellcolor{black}&\cellcolor{black}&\cellcolor{black}&\cellcolor{black}&\cellcolor{black}&\cellcolor{black}&\cellcolor{black}\\\hline
(our solution) IcePwrEst&\cellcolor{black}&\cellcolor{gray}&\cellcolor{black}&\cellcolor{black}&\cellcolor{black}&\cellcolor{gray}&\cellcolor{black}&\cellcolor{black}&\cellcolor{gray}\\\hline
    \end{tabular}
    
    \caption{Feature matrix of power estimation vendor tools}
    \label{tab:featuremtrx}
\end{table}

Awareness of power consumption of an \gls{fpga} design is important for optimization in power or energy constrained use cases.
In this work in progress paper we are presenting the first part of our original work about creating targeted FPGA circuits and measuring them. The second part uses this basis to fit power estimation models. \Cref{tab:featuremtrx} shows the power estimation features offered by different \gls{fpga} vendors and what part of the feature spectrum we want to cover in the second work.
An important goal for us is that all the data and useful tools that we create in this works are going to be published as open source. On the hardware side we are using \gls{fpga}s from the Lattice iCE40 family and the open source tooling based on Yosys and nextpnr that is available for it.

\section{Preliminary}

The previous works that we looked into had two approaches to get power measurement data.
On the one hand they acquired it directly by measurements on hardware \cite{Jevtic2011}, which requires more thought on how to extract certain data. High sensitivity of the measurement system is also important, which can be difficult to achieve on highly clocked systems like MCUs/CPUs \cite{AlShorman2018}.
On the other hand the data can be acquired by utilizing the existing closed source vendor power estimation tools \cite{Verma2018}. Depending on the features of the tool, these directly provide the power of different components.
This second approach requires trust in the accuracy of the vendor provided tools, which can vary to a substantial degree as mentioned in (\cite{Jevtic2011} referencing \cite{Elleouet2006} and \cite{Lee2005}).

\section{Simple Benchmarking}\label{cha:benchmark}
Micro-benchmarks are essential for targeted analysis of certain aspects of FPGA hardware.
These benchmarks can be tedious to create and variate in parameters like component count or placement position.
Thus we decided to design a simplified way of creating such micro-benchmarks by providing a tool, 
which generates said benchmarks in Verilog and an accompanying constraint file from a benchmark definition file.
The micro-benchmarks depend on the target toolchain and hardware, thus requiring specific handling in the generator tool.
In particular we are looking into Lattice iCE40 FPGAs, due to their well supported open source workflow.

\subsection{Introduction}
Micro-benchmarking is an effective mean to analyze certain behavior and parameters of FPGA hardware.
The creation of such benchmarks can be tedious and repetitive, when it is required to place hundreds of identical cells and vary one of their parameters like position or configuration.
The structure of such a generator tool has to be aware of the target toolchain in each step i.e. cell, constraints.
Most of the Verilog code used is toolchain agnostic. Only some special Verilog attributes are toolchain dependent.
The constraint files on the other hand are very different for each vendor and possibly toolchain.
One of the simplifications done to reduce the complexity of the benchmark circuits is to not include an output. This triggered a nextpnr bug\footnote{\label{foot:nextpnr-pr}\href{https://github.com/YosysHQ/nextpnr/pull/944}{https://github.com/YosysHQ/nextpnr/pull/944}} that was fixed by us.

\subsection{Usage and goal}
The micro-benchmark generator is aimed to be used in conjunction with an automated testbench generator and hardware measuring setup.
Our main goal is to accelerate the creation of micro-benchmarks for FPGA hardware analysis by providing a simple tool that generates said benchmarks and its variations. The resulting Verilog files can then be used by other tools also utilizing the definition file to implement testbenches, which can conduct simulations and automated hardware testing.

\subsection{Methodology} 
The basic idea is to instantiate a minimum amount of cells and move them around in the FPGA.
This includes not connecting the output of the component to reduce wires.
Such outputless components are detected by the optimization tool. To prevent them being optimized away it is important to add the Verilog \texttt{keep} attribute.

\begin{figure}[tb]
	\centering
	\includegraphics[width=0.9\linewidth]{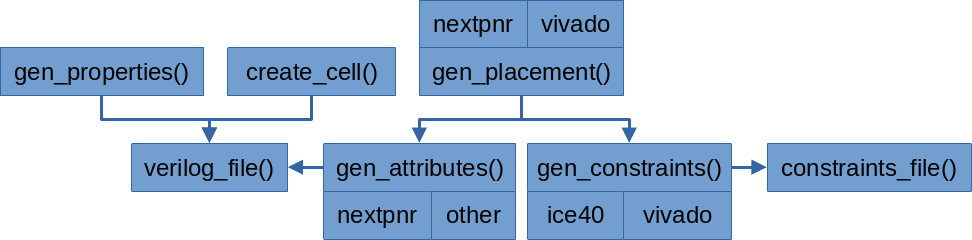}
	\caption{Benchmark generator hierarchy and data flow}
	\label{fig:bench-hiera}
\end{figure}

The various toolchain vendors require different handling of the placement constraints of components. \Cref{fig:bench-hiera} shows the general hierarchy of the functions of the benchmark generator and their data dependencies. 

\subsection{Nextpnr, Yosys -- Lattice}
Nextpnr is a place and route tool that combined with the synthesis tool Yosys supports a fully open source hardware synthesis flow for the Lattice iCE40 FPGA family.
To forcefully place a component with Nextpnr one has to specify the attribute given in \cref{lst:nextpnr}. In example a \gls{lut} is positioned by specifying its X and Y coordinate on the grid of the FPGA and the desired logic cell. The Lattice iCE40up5k has for example 8 logic cells per valid grid position.
\begin{lstlisting}[float,caption={Verilog attribute to place a component into the given logic cell},captionpos=b,label=lst:nextpnr]
	(* BEL="X4/Y4/lc3" *)
\end{lstlisting}

\section{Testbenches}
An accompanying concept to \cref{cha:benchmark} is the automated generation of fitting testbenches. 
The information provided for the benchmark is enough to construct a general stochastic testbench.
This section explains the method used to produce stochastic inputs and how they can be used to test designs.

\subsection{LFSR based testing}
To stimulate the input of a \gls{dut} with stochastic inputs that are not too highly correlated a pseudo random sequence from an \gls{lfsr} can be used.
This testing method is simple to describe for reproduction, because the \gls{lfsr} has only 3 degrees of freedom. 
Those are the polynomial order, the feedback taps and the initial register seed. 
In practice these degrees of freedom can be reduced even further, when only polynomials with maximum length are chosen. 
This limits the sets of feedback taps that can be used for a given polynomial order. 
Lists of polynomial orders and their maximum length tap configurations can be found in most literature about \gls{lfsr}s and also on the internet\footnote{\url{http://users.ece.cmu.edu/~koopman/lfsr/}}.

\subsection{Verilog Testbench}
A Verilog testbench can be utilized to simulate the \gls{dut} with the stochastic inputs that 
will be later applied by the hardware testbenches. The simulation of the testbench can be used 
to acquire an approximate representation of the internal signals of a design running on hardware. 
It is not fully accurate, since no gate or interconnect delays are taken into account. 
The values are rather an lower bound of the internal signal changes. 
These signal changes can be used to calculate activation rates, which are required i.e. for most power estimation models.

\subsection{Microcontroller Testbench}
This testbench is used to examine a design on hardware. It is based on a microcontroller that is connected to 
the \gls{fpga} IOs and the measurement device. The \gls{fpga} contains the \gls{dut} that gets stimulated by 
stochastic signals from the microcontroller. The measurement device receives a trigger signal from 
the microcontroller when the testing begins. The microcontroller itself is controlled by a PC, 
which starts the test runs. The \gls{fpga} is also connected to the PC to allow for exchange of the \gls{dut}.
The data from the measurement device is either directly copied to the PC or indirectly via a data storage depending on the capabilities of the device.

\begin{figure}[tb]
	\centering
	\includegraphics[width=0.8\linewidth]{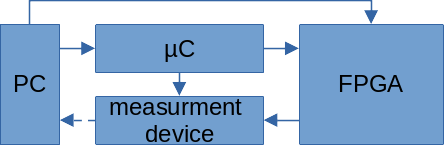}
	\caption{Hierarchy of microcontroller testbench with measurement device}
	\label{fig:uC-testbench-hierachy}
\end{figure}

\section{Experimental Setup}\label{sec:experiment-setup}
Our initial measurement setup \cref{fig:uC-testbench-hierachy} used a USB measurement card with a $12\,\mathrm{Bit}$ \gls{adc}.
At a supply voltage of $5\,\mathrm{V}$ results in at best $\approx 1.2\,\mathrm{mV}$ of quantization step size. The minimal current that can possibly be measured from a $1.5\,\Omega$ shunt resistor is $\approx 800\,\mathrm{\mu A}$.
This turned out to be insufficient measurement accuracy, due to the low power \gls{fpga} used on the iCEBreaker Board. In reality even bigger design targets like the Drystone benchmark\footnote{\url{https://github.com/YosysHQ/picorv32/tree/master/dhrystone}} running on a PicoRV32 core\footnote{\url{https://github.com/YosysHQ/picorv32}} on the iCEBreaker board were in the range of a few quantization steps, which made those measurements unusable.

These problems meant that we had to put more focus on the measuring setup and benchmarking. Our target for useful measurements is around a minimum of $100\,\mathrm{\mu V}$, which would allow currents of $66\,\mathrm{\mu A}$. This seems to be possible to achieve by amplifying the differential voltage of the shunt with an operational amplifier or a dedicated \gls{ic} and measuring the single ended resulting voltage with a accurate \gls{adc}. For this work we pursued a solution with a dedicated \gls{ic}, because such chips have better component matching and less offset errors in addition to very good noise and frequency characteristics. The only real trade off for us was the loss of flexibility, but it was worth the resulting simplicity of the  measurement setup.

\subsection{Current Sense Amplifier Board}\label{subsec:current-sense-amp}

To ease the measurement of small differential voltages on the shunt resistors a current sense amplifier \gls{ic} the INA293B5 has been utilized. The most interesting electrical characteristics of this \gls{ic} are listed in \cref{tab:ina293-datasheet}.

\begin{table}[tb]
	\centering
	\begin{tabular}{|c|c|}
		\hline
		& INA293B5 \\
		\hline
		Gain [V/V] & 500 \\
		\hline
		PSRR [$\mathrm{\mu}$V/V] & 0.1 \\
		\hline
		CMRR [dB] & 140 \\
		\hline
		Bandwidth [kHz] & 900 \\
		\hline
		Voltage noise density [nV/$\surd\mathrm{Hz}$] & 50 \\
		\hline
	\end{tabular}
	\caption{Electrical characteristics of the INA293B5}
	\label{tab:ina293-datasheet}
\end{table}

\subsection{iCEBreaker FPGA Measurement Setup}\label{subsec:icebreaker-measure}
\begin{figure}[tb]
	\centering
	\includegraphics[width=0.5\linewidth]{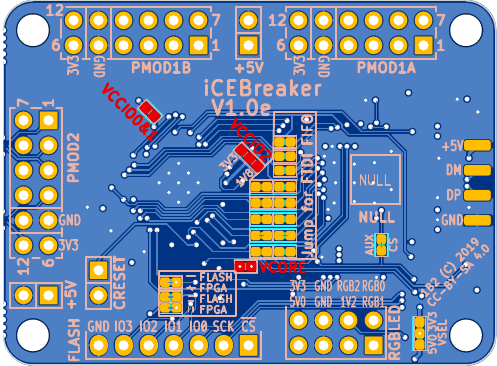}
	\caption{Bottom side of iCEBreaker PCB with position for \textcolor{red}{measurement shunts} marked in \textcolor{red}{red}}
	\label{fig:icebreaker-power-shunts}
\end{figure}

\begin{figure}[tb]
	\centering
	\includegraphics[width=0.5\linewidth,angle=-90]{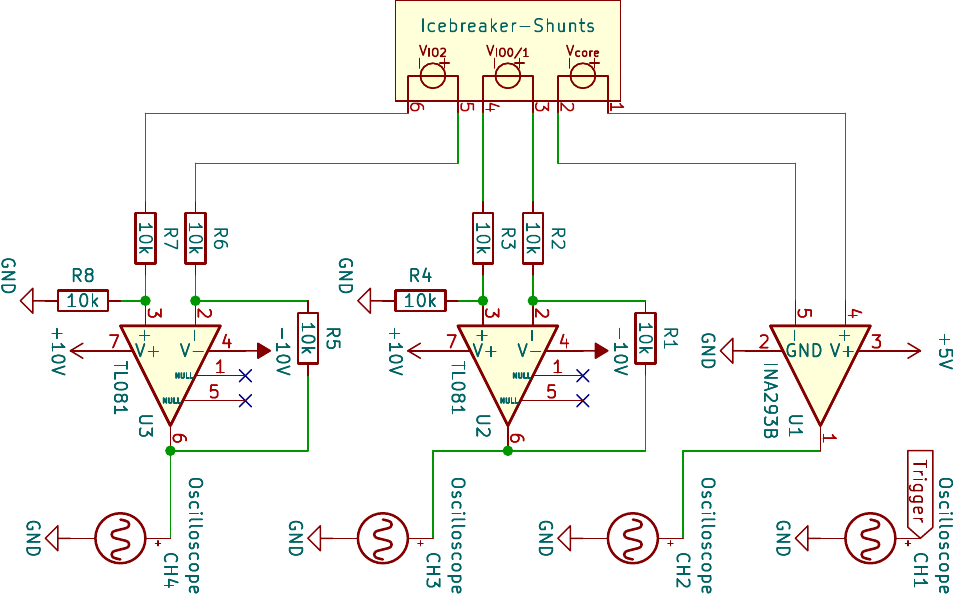}
	\caption{Schematic of the power measurement setup for the iCEBreaker board}
	\label{fig:schematic-power-measurement-setup}
\end{figure}

The iCEBreaker board is using a low power Lattice iCE40UP5k \gls{fpga} \gls{ic}, which has a small amount of logic cells compared to Xilinx \gls{fpga}s. This results in the situation that even the power draw, of designs that fill almost the entire \gls{fpga}, is negligible.
The iCEBreaker PCB has preexisting jumper pads shown in \cref{fig:icebreaker-power-shunts}, which are by design intended to be used for shunt resistors to allow for current measurements of the \gls{fpga}. For our setup $1.5\,\Omega$ resistors have been used.
The setup to measure the power of the board is depicted in \cref{fig:schematic-power-measurement-setup}. The differential voltage of the $V_{\mathrm{core}}$ shunt resistor is getting amplified by the current sense amplifier PCB \cref{subsec:current-sense-amp} and the IO shunt resistors are connected to subtraction circuits due to higher currents, which would clip when amplified. The outputs are then measured with an oscilloscope.

\subsection{Amplifier Prototype Validation}
\begin{figure}[th!]
	\centering
	\includegraphics[width=0.6\linewidth]{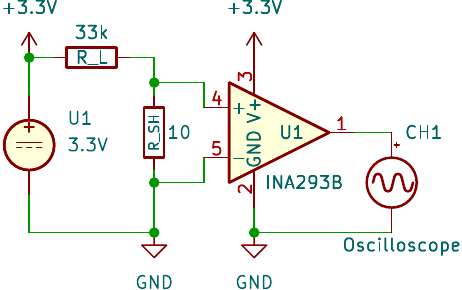}
	\caption{Schematic of the amplifier prototype validation setup}
	\label{sch:amp-proto-val}
\end{figure}

Before the amplifier board could be used for target hardware measurements it had to be validated to confirm it properly functioning. The simple setup for validation is depicted in \cref{sch:amp-proto-val}. The variable values in \cref{tab:hw-measure-amp-proto-spec} have been selected in such a way that a current of approximately $0.1\,\mathrm{mA}$ is measured at the shunt resistor $R_\mathrm{SH}$ resulting in $1\,\mathrm{mV}$. This differential voltage should, if the circuit is working correctly, then get amplified by the INA293 to around $0.5\,\mathrm{V}$.

\begin{table}[tb]
	\centering
	\begin{tabular}{|r|c|c|}
		\hline
		& picked value& measured value  \\
		\hline
		$U1\, [\mathrm{V}]$&$3.3$ & $3.3$ \\
		\hline
		$G\, [\mathrm{V/V}]$&$500$ & -- \\
		\hline
		$R_{\mathrm{L}}\, [\mathrm{k\Omega}]$&$33$ & $31.8$ \\
		\hline
		$R_{\mathrm{SH}}\, [\mathrm{\Omega}]$&$10$ & $10.3$ \\
		\hline
	\end{tabular}
	\caption{Component values of the validation setup}
	\label{tab:hw-measure-amp-proto-spec}
\end{table}
\subsubsection{Measurement Results}
For the measurements it has to be taken into account that the resistance of the connecting points lies in the range of $0.1\textup{--}0.7\,\Omega$, which includes the wire and contact resistance. The resistor tolerance used in the setup was $1\,\%$.
The three amplifiers have been measured one after another. The measurement results of the differential voltage $U_{\mathrm{diff}}$, the amplified output voltage $U_{\mathrm{out}}$ and the calculated amplified voltage $U_{\mathrm{amp\_calc}}$ can be found in \cref{tab:hw-measure-amp-proto-val}. The relative errors of $1.2\textup{--}6.25\,\%$ between the amplified and calculated values seem acceptable when comparing to the uncertainties and variances of the whole setup. The overall behavior of the circuit was correct, which validates its functionality for this DC test.
\begin{table}[tb]
	\centering
	\begin{tabular}{|r|c|c|c|}
		\hline
		& Amp1 & Amp2 & Amp3\\
		\hline
		$U_{\mathrm{out}}\, [\mathrm{V}]$&0.415  & 0.417 & 0.416\\
		\hline
		$U_{\mathrm{diff}}\, [\mathrm{mV}]$& 0.82 & 0.8  & 0.78\\
		\hline
		$U_{\mathrm{amp\_calc}}\, [\mathrm{V}]$& 0.41 & 0.4  & 0.39\\
		\hline
		$e_{\mathrm{rel}}\, [\mathrm{\%}]$& 1.20 & 4.08  & 6.25\\
		\hline
	\end{tabular}
	\caption{Measurements from the validation setup}
	\label{tab:hw-measure-amp-proto-val}
\end{table}

\section{iCEBreaker FPGA -- Measurements}
The measurements of the Lattice iCE40UP5k \gls{fpga} are of particular interest to us for analyzing its internals. To gather information about the properties of the internals of the \gls{fpga}, different circuits and measurements have been used. The following sections go over the different circuits and their measurement results.
\subsection{LUT4 -- Results} \label{subsec:lut-lfsr}
One of the basic internal building blocks of the \gls{fpga} is the \gls{lut} with 4 inputs and one output, which was measured in this section.
The circuit in this setup consists of only \gls{lut}, that get instantiated by the benchmark generator in \cref{cha:benchmark}. The inputs of all \gls{lut}s get connected to 4 inputs of the \gls{fpga} in parallel. This allows to control all \gls{lut}s by only using 4 inputs. The idea behind connecting all \gls{lut}s in parallel is to be able to measure their power in an additive way. This should result in a proportional increase of power in regards to the number of \gls{lut}s used. For example reducing the \gls{lut} amount from $5\,\mathrm{k}$ to $1\,\mathrm{k}$ would result in approximately a fifth of the power after correcting for a constant offset value.

\begin{figure}[tb]
	\centering
	\subfloat[$5\,\mathrm{k}$ LUT4]
	{\label{oszi:5klut}\includegraphics[width=0.5\linewidth]{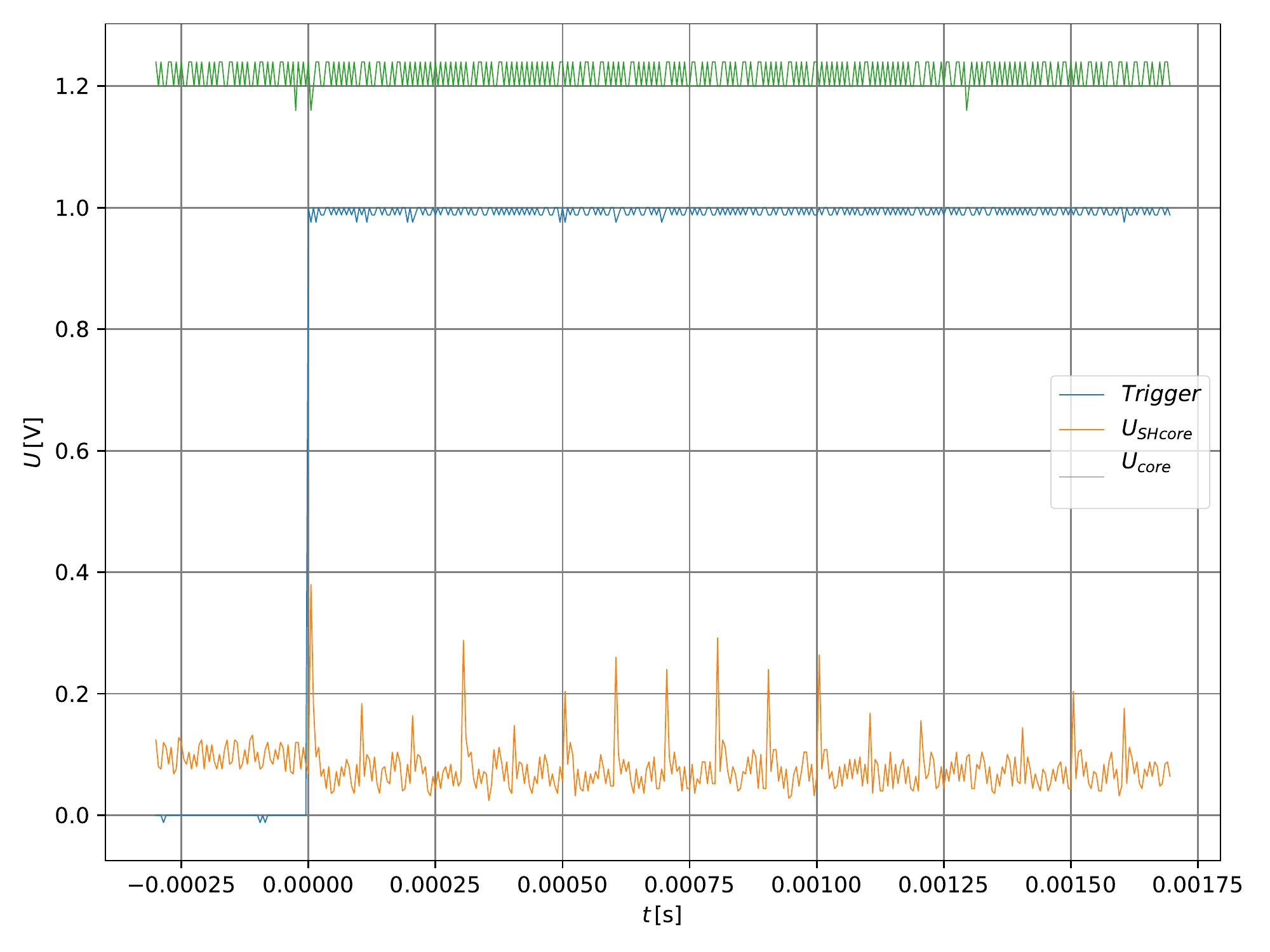}}
	\subfloat[$1\,\mathrm{k}$ LUT4]
	{\label{oszi:1klut}\includegraphics[width=0.5\linewidth]{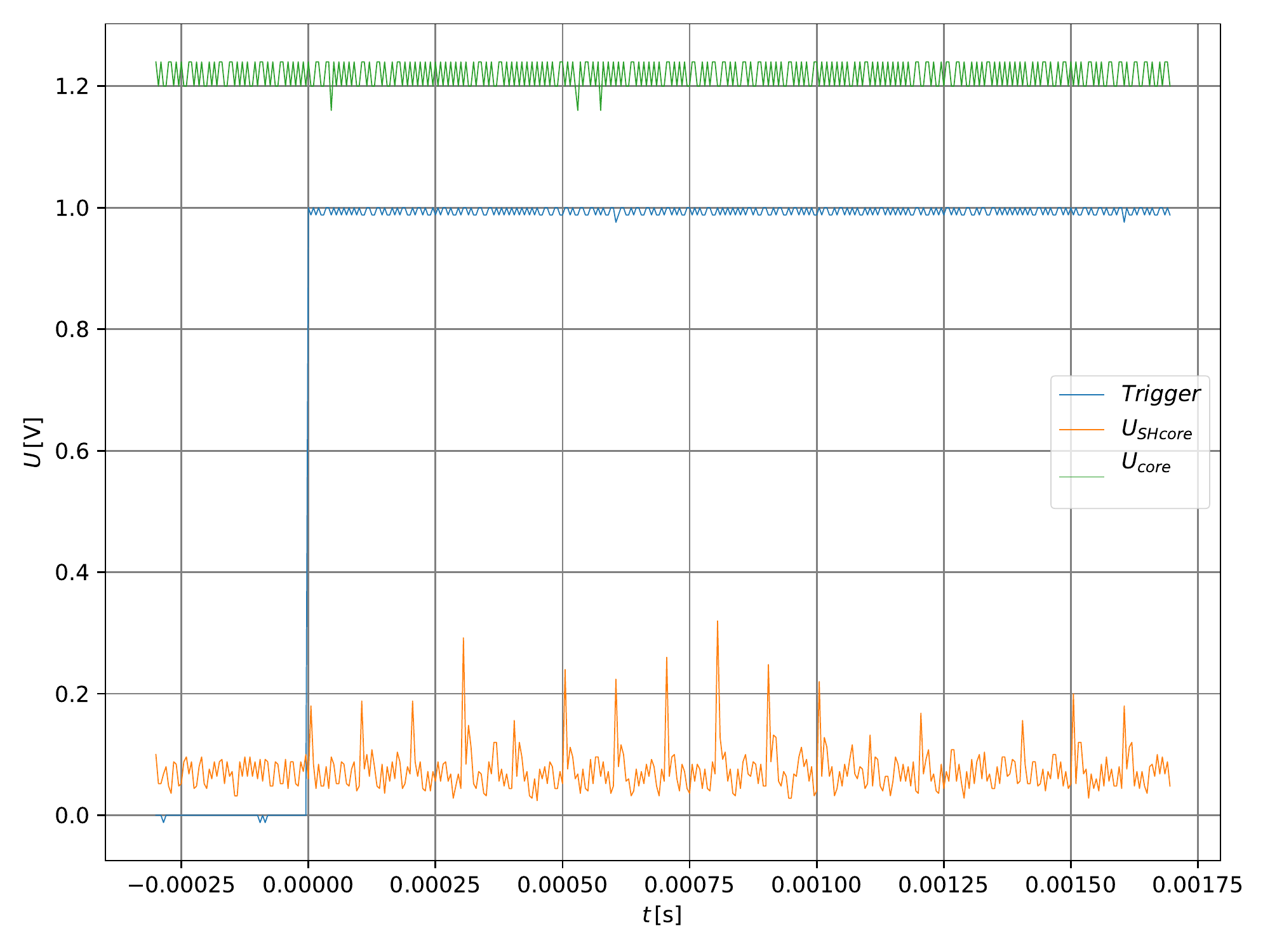}}
	
	\caption{Measurements of LUTs stimulated by 4Bit-\gls{lfsr}}
	\label{fig:lfsr4lut}
\end{figure}

\begin{figure}[tb]
	\centering
	\subfloat[stimulated by 1 output of 4Bit-\gls{lfsr}]
	{\label{oszi:1lut}\includegraphics[width=0.5\linewidth]{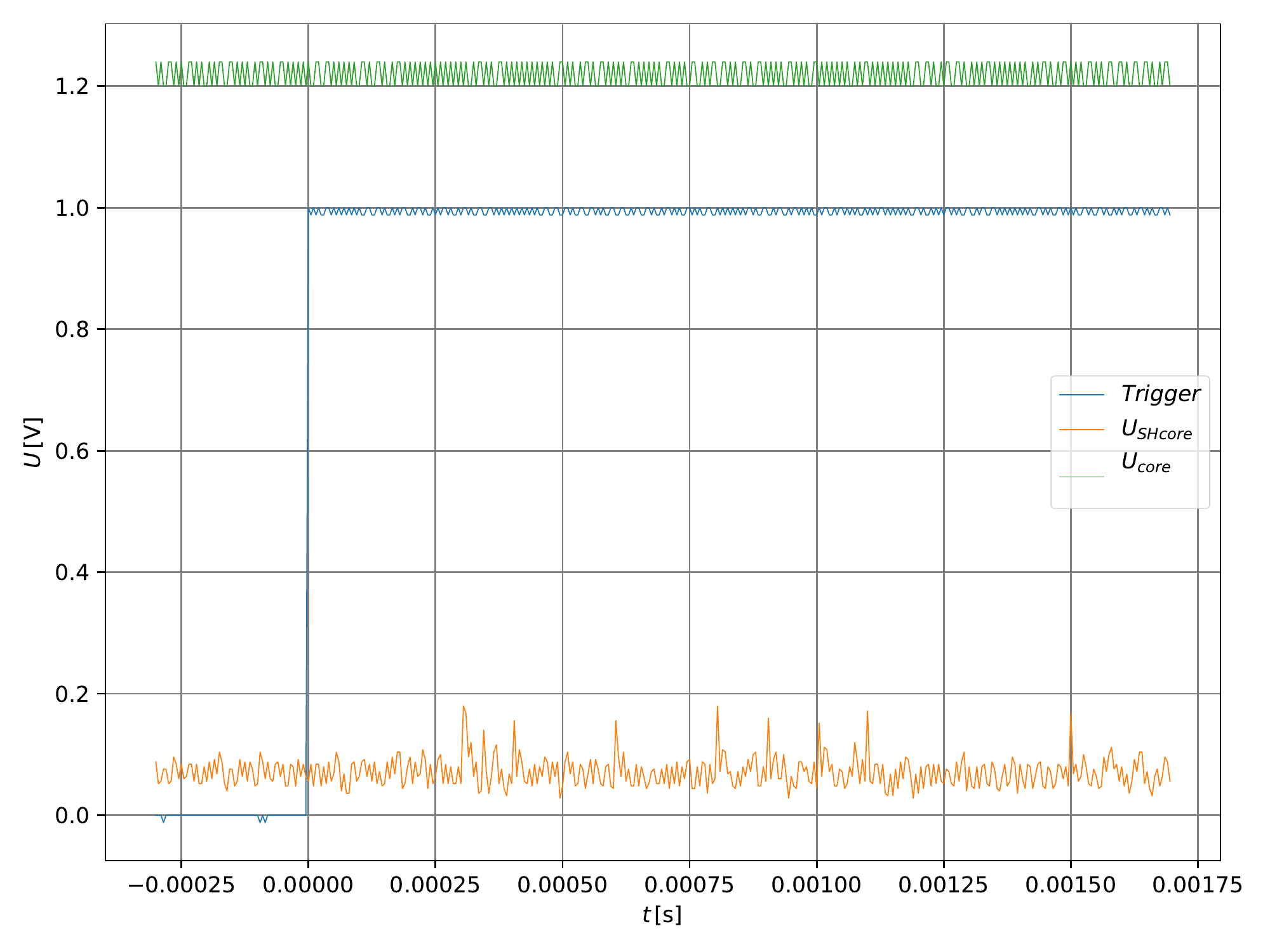}}
	\subfloat[not
	stimulated]{\label{oszi:1lutnoio}\includegraphics[width=0.5\linewidth]{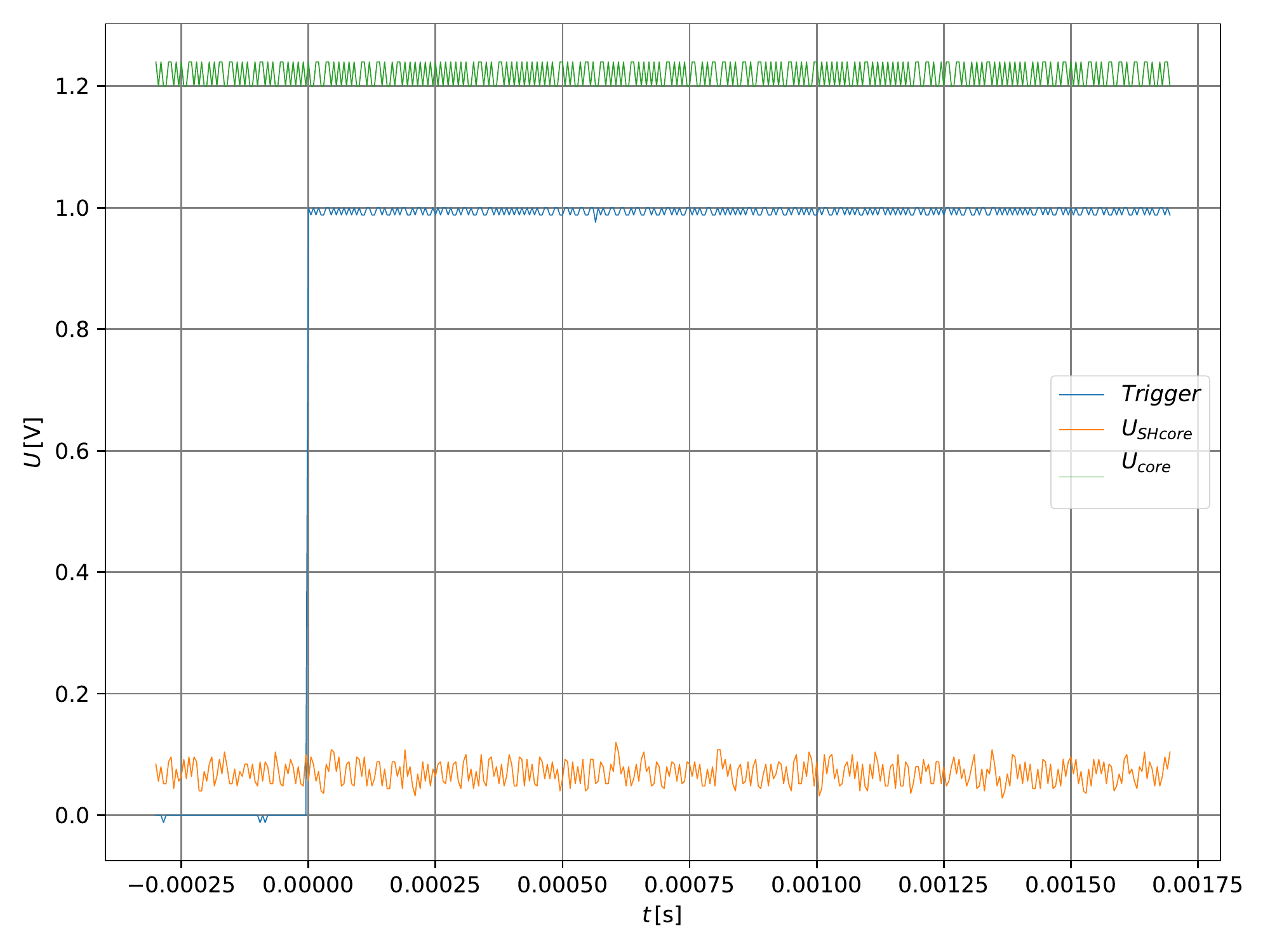}}
	
	\caption{Measurements of 1 LUT4}
	\label{fig:lfsr1lut}
\end{figure}

\Cref{fig:lfsr4lut} contains measurements of the iCEBreaker board with $5\,\mathrm{k}$ and $1\,\mathrm{k}$ \gls{lut} instantiated on the \gls{fpga} respectively. These inputs then got stimulated by a maximum length 4Bit-\gls{lfsr} every $100\,\mathrm{\mu{}s}$ after the rising trigger signal. When looking at \cref{fig:lfsr4lut} one can see the voltage spikes in a regular pattern. This pattern matches the \gls{lfsr} changing values and one can also see its period of 15. Comparing the three graphs does not bring the desired result of proportional dependency of the power to the number of \gls{lut}. This means the voltage spikes on the shunt are the result of something else.
For further investigation \cref{fig:lfsr1lut} has been created. The setup for these measurements was one instantiated \gls{lut} and only one input connected to one bit of the \gls{lfsr}. The purpose of this measurement was to investigate how the power relates to the connected signals and their state.

Comparing the measurements from \cref{fig:lfsr4lut} and \cref{fig:lfsr1lut} shows that the voltage on the shunt resistor does not depend on the number of \gls{lut} at all, due to there being no real difference in the amplitude of the voltage spike. On the other hand the comparison allows the conclusion that the amplitude of the voltage spike depends on the amount of inputs in high state. In \cref{fig:lfsr1lut} the voltage spikes occur only sparsely due to the \gls{lfsr} still being of period 15 and only one bit being connected to the \gls{lut}.

\subsection{Ring-oscillator}\label{subsec:ring-osc}

The next circuit to test for power measurements is a simple ring oscillator.
This can be constructed by daisy chaining an uneven number of \texttt{NOT}-gates together and connecting the output of the last gate with the input of the first. Such a construct is used on \gls{fpga}s for example to generate random numbers.

As a first attempt to instantiate a ring oscillator the Verilog code in \cref{lst:ringoscillator-not} has been used. This code should generate a ring oscillator that can be turned on and off with an \gls{fpga} input via the \texttt{AND}-gate. The \texttt{keep} attribute should prevent the synthesis tool from optimizing the gates away.

\begin{figure}[tb]
	\centering
	\includegraphics[width=0.9\linewidth]{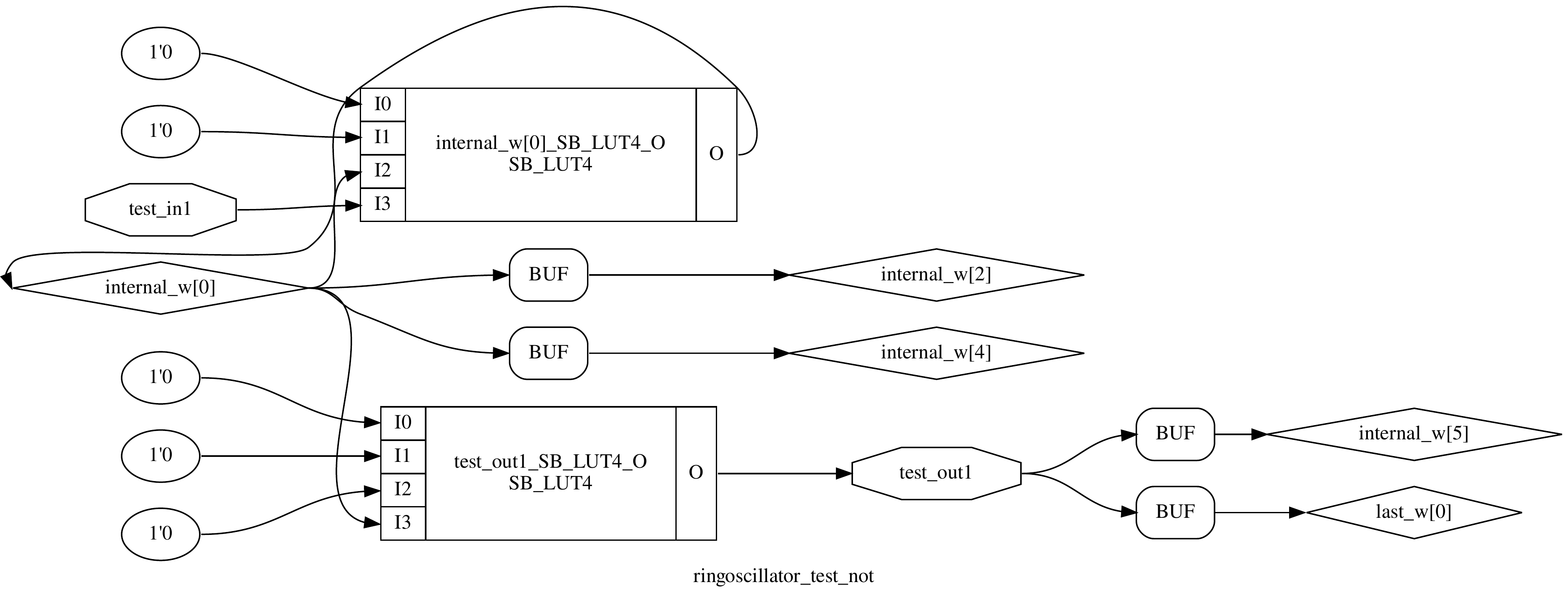}
	\caption{Schematic of the synthesized \texttt{not} based ring oscillator}
	\label{fig:schematic-ringoscillator-not}
\end{figure}

After a few synthesis steps in Yosys the resulting circuit in \cref{fig:schematic-ringoscillator-not} shows that the attempt of using \texttt{keep} was futile and that most gates got optimized away. The reason for the optimization kicking in is due to the translation of the \texttt{NOT} and \texttt{AND}-gates to \gls{lut}, which the \texttt{keep} attribute did not propagate to. This means that a more elaborate way to instantiate the ring oscillator has to be used.

\begin{lstlisting}[float,language=Verilog,caption={Example code of Verilog high level
	ringoscillator implementation},captionpos=b,label=lst:ringoscillator-not]
module ringoscillator_test_not(
    input [0:0] test_in1,
    output [0:0]test_out1
);

wire internal_w[5:0];
wire last_w[0:0];

assign test_out1[0] = last_w[0];
(* keep  *)
and(internal_w[0],test_in1[0],last_w[0]); //and(Y,A,B) Y = A&B
(* keep  *)
not(internal_w[1],internal_w[0]); //not(Y,A) Y = ~A
(* keep  *)
not(internal_w[2],internal_w[1]);
(* keep  *)
not(internal_w[3],internal_w[2]);
(* keep  *)
not(internal_w[4],internal_w[3]);
(* keep  *)
not(internal_w[5],internal_w[4]);
assign last_w[0] = internal_w[5];
endmodule
\end{lstlisting}

\begin{figure}[tb]
	\centering
	\includegraphics[width=0.9\linewidth]{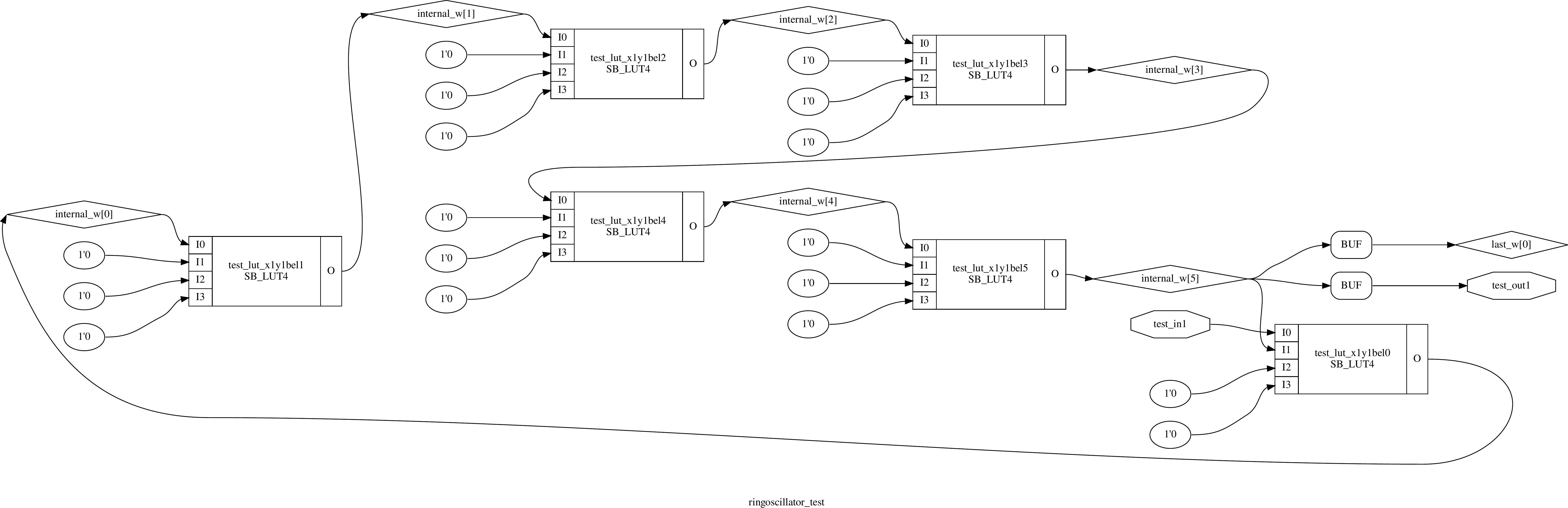}
	\caption{Schematic of the synthesized \gls{lut} based ring oscillator}
	\label{fig:schematic-ringoscillator-lut}
\end{figure}

Utilizing the tricks learned from the benchmark generator we can approach the issue from the further down the synthesis steps.
The code in \cref{lst:ringoscillator-lut} shows a functional equivalent implementation of the ring oscillator as described in \cref{lst:ringoscillator-not}, but this time the general gates have been replaced with the iCE40 specific \gls{lut}. \Cref{fig:schematic-ringoscillator-lut} shows that this time the synthesis tool did not optimize the intermediate gates away. This means with this approach we can generate the desired ring oscillators that we want to measure on \gls{fpga} hardware.

\begin{lstlisting}[float,language=Verilog,caption={Example code of Verilog low level
	ringoscillator implementation},captionpos=b,label=lst:ringoscillator-lut]
module ringoscillator_test(
    input test_in1,
    output test_out1
);

parameter [15:0] not_cfg= 16'b0000_0000_0000_0001;
parameter [15:0] and_cfg= 16'b0000_0000_0000_1000;

wire internal_w[5:0];
wire last_w[0:0];

assign test_out1 = last_w[0];

(* keep  *) (* BEL="X1/Y1/lc0" *)
SB_LUT4 #(.LUT_INIT(and_cfg)) test_lut_x1y1bel0(.I0(test_in1),.I1(last_w[0]),.I2(1'b0),.I3(1'b0),.O(internal_w[0])); //and
(* keep  *) (* BEL="X1/Y1/lc1" *)
SB_LUT4 #(.LUT_INIT(not_cfg)) test_lut_x1y1bel1(.I0(internal_w[0]),.I1(1'b0),.I2(1'b0),.I3(1'b0),.O(internal_w[1])); //not
(* keep  *) (* BEL="X1/Y1/lc2" *)
SB_LUT4 #(.LUT_INIT(not_cfg)) test_lut_x1y1bel2(.I0(internal_w[1]),.I1(1'b0),.I2(1'b0),.I3(1'b0),.O(internal_w[2])); //not
(* keep  *) (* BEL="X1/Y1/lc3" *)
SB_LUT4 #(.LUT_INIT(not_cfg)) test_lut_x1y1bel3(.I0(internal_w[2]),.I1(1'b0),.I2(1'b0),.I3(1'b0),.O(internal_w[3])); //not
(* keep  *) (* BEL="X1/Y1/lc4" *)
SB_LUT4 #(.LUT_INIT(not_cfg)) test_lut_x1y1bel4(.I0(internal_w[3]),.I1(1'b0),.I2(1'b0),.I3(1'b0),.O(internal_w[4])); //not
(* keep  *) (* BEL="X1/Y1/lc5" *)
SB_LUT4 #(.LUT_INIT(not_cfg)) test_lut_x1y1bel5(.I0(internal_w[4]),.I1(1'b0),.I2(1'b0),.I3(1'b0),.O(internal_w[5])); //not
assign last_w[0] = internal_w[5];
endmodule
\end{lstlisting}

\subsubsection{iVerilog iCE40 simulation with delays}\label{subsubsec:iverilog-sim-delay}

Having a working simulation environment to work with is useful to evaluate the behavior of designs beforehand. This is especially important for low level implementations on \gls{fpga} logic blocks and high frequency circuits like ring oscillators. For the first type a zero delay simulation is good enough, but the latter requires approximate timings of the \gls{fpga}. A high frequency ring oscillator can cause the destruction of the \gls{fpga} by overloading the internal paths due to high frequency switching. Synthesis tools might detect such high frequency paths and prevent them, but designers can choose to implement the oscillator anyways. To prevent the destruction it is important to know the approximate oscillation frequency, which can be determined from simulation that includes the gate delays. The frequency will be an upper bound to the real frequency, due to path delays being missing in this step.

For Verilog simulation the open source tool \gls{iverilog} can be used. Verilog synthesized or tailored to the iCE40 architecture can be simulated with \gls{iverilog} by including the simulation library from Yosys.
It is to note the the Verilog path delay definitions in Yosys' simulation library do not comply to the Verilog standard and \gls{iverilog} errors out when parsing them. This is fixed by our pull request\footnote{\url{https://github.com/YosysHQ/yosys/pull/3542}} to the library.

\Cref{lst:iverilog-sim} shows an example on how to simulate an iCE40 synthesized design with \gls{iverilog}. The argument \texttt{-gspecify} is necessary to enable the usage of Verilog \texttt{specify} blocks, which contain the path delay definitions. With the flag \texttt{-D} the corresponding Verilog define variables are being set. Yosys' simulation library contains timing definitions for the different variants of the iCE40 architecture. To select the low power variant the variable \texttt{'ICE40\_LP=1'} has to be defined.

\begin{lstlisting}[float,language=bash,caption={Example code to run a simulation including path delays with \gls{iverilog}},captionpos=b,label=lst:iverilog-sim]
	iverilog \
	-gspecify \
	-D 'VCDFILE="simulation_trace.vcd"' \
	-D 'NO_ICE40_DEFAULT_ASSIGNMENTS=1' \
	-D 'ICE40_LP=1' \
	-o simulation_output \
	$(yosys-config --datdir/ice40/cells_sim.v)) \
	test_ice40.v
	vvp simulation_output
\end{lstlisting}

\subsubsection{Results}
For the following results use a total number of approximately 3000 \texttt{NOT} gates has been instantiated on the iCEBreaker board. These gates are all chained together in the same fashion as shown in \cref{lst:ringoscillator-lut} for the first ring oscillator $O1$. The second the design $O2$ consists of two chains of half the length of $O1$ with double the oscillation frequency. The third the design $O4$ consists of half the length of $O2$ with double the chains, which results in two times the frequency.

\begin{figure}[tb]
	\centering
	\includegraphics[width=0.9\linewidth]{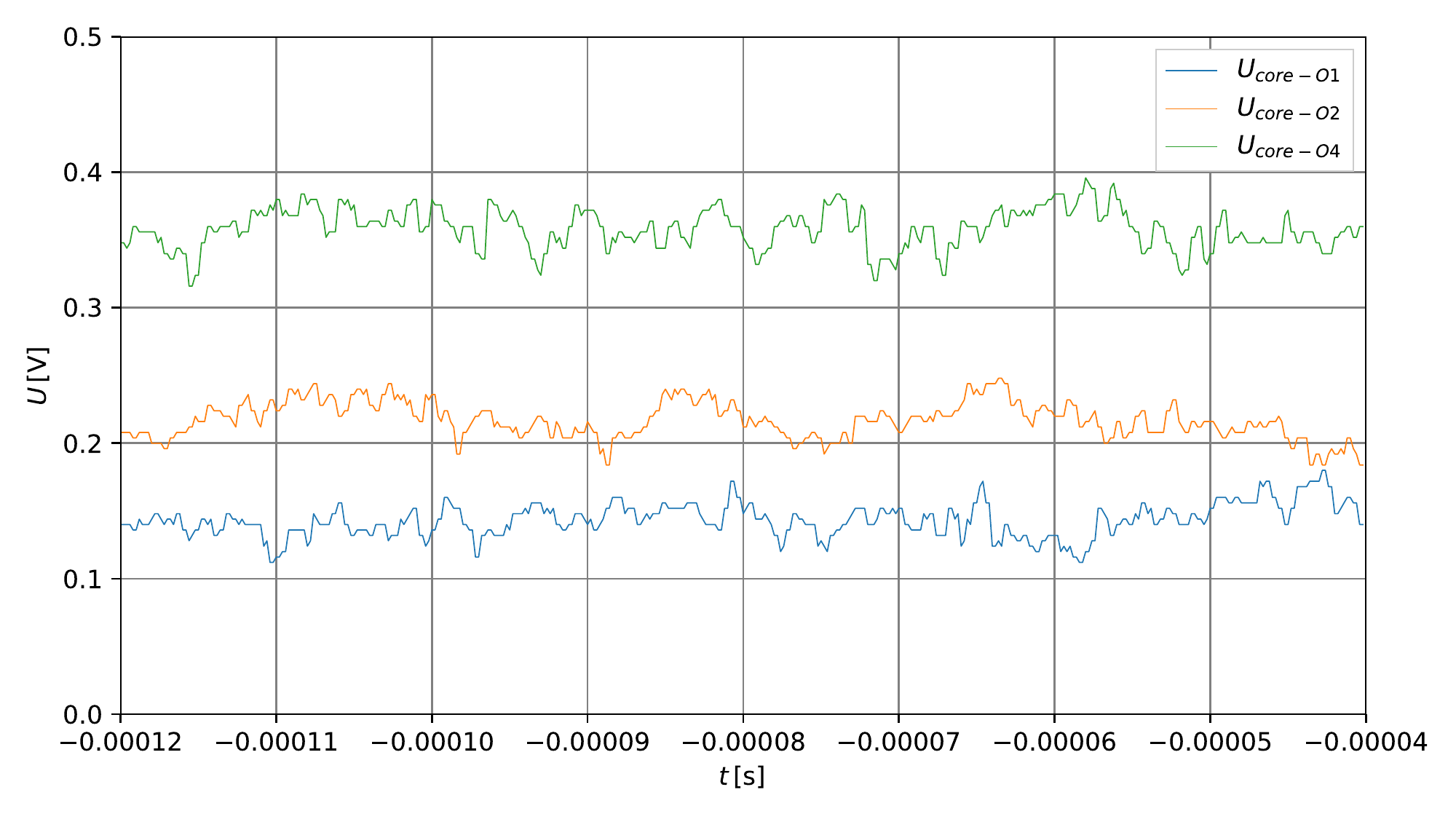}
	\caption{Differential voltage on the $V_{\mathrm{core}}$ shunt resistor of three different ring oscillators}
	\label{oszi:ring-comb}
\end{figure}

\begin{figure}[tb]
	\centering
	\includegraphics[width=0.9\linewidth]{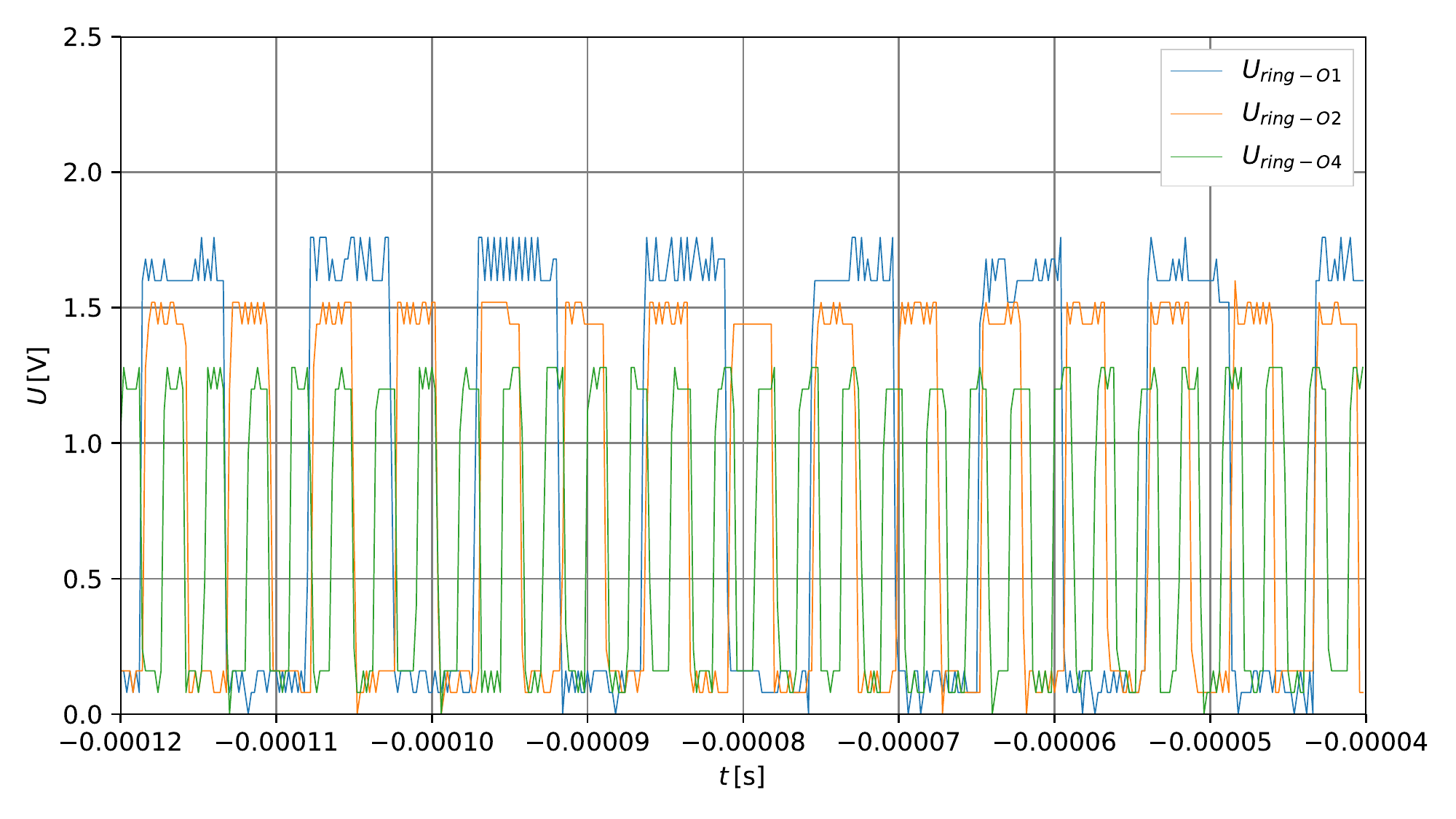}
	\caption{Outputs of the three ring oscillators used}
	\label{oszi:ring-osc}
\end{figure}

\Cref{oszi:ring-osc} shows the output of the oscillating signal of one chain in each of the three designs. One can clearly observe the doubling of the oscillator frequency between each of the designs. This confirms the oscillator designs properly working and scaling. Now the most interesting measurements are found in \cref{oszi:ring-comb}, which shows an increasing voltage on the $V_{core}$ shunt that corresponds to the core current. Assuming an approximately constant core voltage it is trivial to calculate the dynamic power used based on the following simple model.

\section{Conclusion}
Initial results of \cref{sec:experiment-setup} showed that measuring low power \gls{fpga}s is not a simple task. Solely an improved measurement setup compared to the USB measurement card was not sufficient to get sensible results as depicted in \cref{subsec:lut-lfsr}. The influence of the external input signals was too severe and the low stimulation frequency didn't achieve significant power draw changes. This meant that the results can not be used for our intended purpose.
The latter results in \cref{subsec:ring-osc} on the other hand show a significant difference between the three designs, which provides strong evidence that the characteristics of the ring oscillator alone resulted in this change. Due to the ring oscillator being solely constructed out of iCE40 \gls{lut}s means that the measured data contains information about their power characteristics.

\section{Future Outlook}
Further analysis of the measurements in \cref{subsec:ring-osc} has to be done to provide stronger evidence for their plausibility.
Based on these results we will expand the set of test circuits to acquire more data.
This will enable us to design appropriate estimation models of the analyzed Lattice hardware.
These models will then be fitted with the data to allow for power estimation, which will then be checked with new test circuits against hardware measurements.


{
\footnotesize
\begin{thebibliography}{00}
\tiny
\bibitem{Jevtic2011} Ruzica Jevtic and Carlos Carreras, ``
Power Measurement Methodology for FPGA Devices``,
IEEE Transactions on Instrumentation and Measurement, Institute of Electrical and Electronics Engineers (IEEE), 2011, 60, 237-247.
\bibitem{Verma2018} Gaurav Verma and Vijay Khare and Manish Kumar,
``More Precise FPGA Power Estimation and Validation Tool (FPEV\_Tool) for Low Power Applications``,
Wireless Personal Communications, Springer Science and Business Media LLC, 2018, 106, 2237-2246
\bibitem{AlShorman2018} Mohammad Y Al-Shorman and Majd M Al-Kofahi and Osameh M Al-Kofahi,
``A practical microwatt-meter for electrical energy measurement in programmable devices 
Measurement and Control``, SAGE Publications, 2018, 51, 383-395
\bibitem{Elleouet2006} David Ell{\'{e}}ouet and Yannig Savary and Nathalie Julien,
``An FPGA Power Aware Design Flow``,
Lecture Notes in Computer Science, Springer Berlin Heidelberg, 2006, 415-424
\bibitem{Lee2005} Hyung Gyu Lee and Kyungsoo Lee and Yongseok Choi and Naehyuck Chang, ``Cycle-Accurate Energy Measurement and Characterization of FPGAs``,
Analog Integrated Circuits and Signal Processing, Springer Science and Business Media LLC, 2005, 42, 239-251
\end{thebibliography}}
\end{document}